\documentclass[tighten,twocolumn,apj]{likeapj}
\pdfoutput=1 

\hypersetup{linkcolor=blue,citecolor=blue,filecolor=cyan,urlcolor=blue,bookmarksnumbered=true,bookmarksopen=true}
\usepackage[all]{hypcap}

\usepackage{etoolbox}
\makeatletter

\patchcmd{\NAT@citex}
  {\@citea\NAT@hyper@{%
     \NAT@nmfmt{\NAT@nm}%
     \hyper@natlinkbreak{\NAT@aysep\NAT@spacechar}{\@citeb\@extra@b@citeb}%
     \NAT@date}}
  {\@citea\NAT@nmfmt{\NAT@nm}%
   \NAT@aysep\NAT@spacechar\NAT@hyper@{\NAT@date}}{}{}

\patchcmd{\NAT@citex}
  {\@citea\NAT@hyper@{%
     \NAT@nmfmt{\NAT@nm}%
     \hyper@natlinkbreak{\NAT@spacechar\NAT@@open\if*#1*\else#1\NAT@spacechar\fi}%
       {\@citeb\@extra@b@citeb}%
     \NAT@date}}
  {\@citea\NAT@nmfmt{\NAT@nm}%
   \NAT@spacechar\NAT@@open\if*#1*\else#1\NAT@spacechar\fi\NAT@hyper@{\NAT@date}}
  {}{}

  \makeatother
\usepackage{mathptmx}

\usepackage{amsmath}
\usepackage{natbib}
\newcommand{\hm}{$h^{-1}$}

\newcommand{\nmin}{$N_{\rm min}$}
\newcommand{\scr}{Section}

\received{2017 April 26}
\revised{2019 January 23}
\accepted{2019 January 25}
\published{2019 March 12}
\shorttitle{Compact Groups Through Cosmic Time}

\shortauthors{Wiens et al.}

\begin{document}

\title{The Occurrence of Compact Groups of \\Galaxies Through Cosmic Time}

\author{Christopher D. Wiens}
\affiliation{Department of Astronomy, University of Virginia, P.O. Box 400325, Charlottesville, VA 22904-4325, USA}

\author{Trey V. Wenger}
\affiliation{Department of Astronomy, University of Virginia, P.O. Box 400325, Charlottesville, VA 22904-4325, USA}
\affiliation{National Radio Astronomy Observatory, 520 Edgemont Road, Charlottesville, VA 22903, USA}

\author{Panayiotis Tzanavaris}
\affiliation{Astrophysics Science Division, Laboratory for X-ray Astrophysics, NASA/Goddard Space Flight Center, Mail Code 662, Greenbelt, MD 20771, USA}
\affiliation{CRESST, University of Maryland Baltimore County, 1000 Hilltop Circle, Baltimore, MD 21250, USA}

\author{Kelsey E. Johnson}
\affiliation{Department of Astronomy, University of Virginia, P.O. Box 400325, Charlottesville, VA 22904-4325, USA}
\affiliation{National Radio Astronomy Observatory, 520 Edgemont Road, Charlottesville, VA 22903, USA}

\author{S.~C.~Gallagher}
\affil{Department of Physics and Astronomy, University of Western Ontario, London, ON N6A 3K7, Canada}
\affil{Centre for Planetary and Space Exploration, University of Western Ontario, London, ON N6A 5B9, Canada}
\affil{Rotman Institute of Philosophy, University of Western Ontario, London, ON N6A 5B9, Canada}
\affil{Canadian Space Agency, Saint-Hubert, QC J3Y 8Y9, Canada}

\author{Liting Xiao}
\affiliation{Department of Astronomy, University of Virginia, P.O. Box 400325, Charlottesville, VA 22904-4325, USA}
\affiliation{Division of Physics, Mathematics and Astronomy, California Institute of Technology, Pasadena, CA 91125, USA}
\begin{abstract}
We use the outputs of a semianalytical model of galaxy
formation run on the Millennium Simulation to investigate the prevalence of 3D compact groups (CGs) of galaxies from $z = 11$ to 0. Our publicly available code identifies CGs 
%
%
%
using the 3D galaxy number density, the mass ratio of secondary$+$tertiary to the primary member, mass density in a surrounding shell, the relative velocities of candidate CG members, and a minimum CG membership of three.
%
We adopt ``default'' values for the first three criteria, representing the observed population of Hickson CGs at $z = 0$. The percentage of nondwarf galaxies ($M > 5 \times 10^{8}h^{-1}\ M_{\odot}$) in CGs peaks near $z \sim 2$ for the default set and in the range of $z \sim 1$--3 for other parameter sets. This percentage declines rapidly at higher redshifts ($z \gtrsim 4$), consistent with the galaxy population as a whole being dominated by low-mass galaxies excluded from this analysis. According to the most liberal criteria, $\lesssim$3\% of nondwarf galaxies are members of CGs at the redshift where the CG population peaks. Our default criteria result in a population of CGs at $z < 0.03$ with number densities and sizes consistent with Hickson CGs.  Tracking identified CG galaxies and merger products to $z = 0$, we find that $\lesssim$16\% of nondwarf galaxies have been CG members at some point in their history. Intriguingly, the great majority (96\%) of $z = 2$ CGs have merged to a single galaxy by $z= 0$. There is a discrepancy in the velocity dispersions of Millennium Simulation CGs compared to those in observed CGs, which remains unresolved.
    \end{abstract}

\hspace{1cm}

\section{Introduction}
Determining which physical mechanisms dominate the processing of gas in galaxies is at the core of understanding galaxy evolution over cosmic time. 
With high apparent galaxy number densities and relatively low velocity dispersions \citep{hickson92,mamon92}, compact groups of galaxies 
(CGs) appear to be the ideal environment to investigate how gas processing is impacted by relatively strong interactions between multiple galaxies {\it simultaneously}.  

Recent work has demonstrated that the CG environment has an impact on galaxy evolution that is not seen in several other environments, including field galaxies, isolated pairwise mergers, and galaxy clusters \citep{johnson07, tzanavaris10, walker10, walker12, lenkic16,  zucker16}. Specifically, galaxies that reside in CGs exhibit a ``canyon'' in infrared color space that implies a rapid transition between relatively actively star-forming and quiescent systems.  Curiously, it should be noted that this transition region in infrared color space corresponds to the optical ``red sequence'' rather than the ``green valley'' \citep{walker13,zucker16}, indicating that a distinct evolutionary process is taking place in CGs. In other words, the infrared transition region seen in CGs, is not seen in comparison samples and should not be confused with the optical green valley.

This is further corroborated by a bimodality in star formation rates normalized by stellar mass (specific star formation rates, [sSFRs]), which is particular to the CG environment \citep{tzanavaris10,lenkic16}, although some bimodal behavior has been reported in loose groups as well \citep{wetzel12}. Bimodal star formation suppression was also
reported by \citet{alatalo15}, who studied warm H$_2$ gas in CGs \citep[see also][]{cluver13}. Peculiar sSFR behavior was also reported by
\citet{bitsakis10, bitsakis11}, who found that late-type galaxies in CGs have systematically low sSFRs. \citet{lisenfeld17} found that some CG galaxies appear to have a lower star-formation efficiency (SFR/$M_{\rm H_2}$); however, star-forming CG galaxies do lie on the star-forming “main sequence,” consistent with galaxies in other environments \citep{lenkic16}.

A number of other works have shown that galaxy evolution is impacted by the CG environment.
\citet{proctor04}, \citet{mendesdeoliveira05}, and \citet{coenda15} found that CG galaxies tend to be older {(in terms of the average age of their stellar populations)} than galaxies in other environments.
\citet{coenda12} found a significantly larger fraction of red and early-type galaxies in CGs, as compared to loose groups, while
\citet{martinez13} established that brightest group galaxies in CGs are
brighter, more massive, larger, redder, and more frequently classified
as elliptical compared to their counterparts in loose groups.
\citet{coenda15} found that
CGs include a late-type population with markedly reduced sSFRs compared to loose groups and field populations.
The fraction of quiescent galaxies (i.e., not actively star-forming, independent of the average age of the stellar population) in CGs is higher than in the field or loose-group population \citep{coenda15,lenkic16}.

\citet{farhang17} compared CGs to fossil groups in the Millennium Simulation, finding that some, but not all, CGs eventually turn into fossil systems. They stressed that CGs appear to be a distinct group environment.

The fact that CGs show distinct features and/or behavior 
when compared to other environments suggests that the CG environment is
``doing {\it something}'' that impacts galaxy evolution in a unique way, not prevalent in other environments. In turn, this may be linked to the high, present or past, interaction activity experienced by galaxies in these systems \citep[see, e.g., the detailed results on interactions in CG systems in][]{plana98,mendesdeoliveira98,mendesdeoliveira03,amram04,torres-flores09,torres-flores10,torres-flores14}.

The importance of understanding how galaxy evolution is impacted by the group environment is underscored by the fact that most galaxies spend the majority of their time in groups of some kind \citep[e.g.][and references therein]{mulchaey00,karachentsev05}. However, the fraction of galaxies that have been part of a CG over cosmic time is unclear, and therefore the total impact of the CG environment on galaxy evolution throughout the universe has not been well constrained. Catalogs of CGs are restricted to the relatively nearby universe (e.g., \citealt{hickson92, barton96, lee04, deng07, mcconnachie09, diaz-gimenez12}; but see \citealt{pompei12}), and even in these cases CGs can only be confirmed if velocity information is available for the constituent galaxies.  For example, the Redshift Survey CG catalog \citep{barton96} has a magnitude limit of $m_B < 15.5$, only reaching a redshift of $z\lesssim 0.03$. Advances in galaxy simulations over the past decade now enable us to explore the prevalence of CGs out to arbitrarily high redshifts.

\subsection{Simulations}
\setcounter{footnote}{9}

As it is not possible to observe a sample of CGs spanning all redshifts, the only available path forward is to utilize cosmic simulations of galaxy evolution. The 
Millennium Simulation \citep{springel05,lemson06} 
provides a tool to study the history of CGs in the universe. It used \(2160^3\) particles with mass \(M = 8.6\times10^8 h^{-1} M_\odot\) to trace the evolution of a comoving cube with side length \(500 \, h^{-1}\) Mpc, where $h$ is the Hubble constant parameterization such that $H_0=100 \, h$ \text{km s\(^{-1}\) Mpc\(^{-1}\)}. The Millennium Simulation used a \(\Lambda\)CDM model with cosmological parameters \(\Omega_M = 0.25\), \(\Omega_b = 0.045\), \(\Omega_\Lambda = 0.75\), and \(h = 0.73\), consistent with observational results from the Carnegie Hubble Program \citep{freedman2012} and the WMAP \citep{hinshaw13} mission.\footnote
{In this paper all distances and masses are scaled to $h=1$.}

The Millennium Simulation used only dark matter particles to which associated properties were later assigned. Dark matter halos were identified using a friends-of-friends algorithm as described in \citet{springel05}.  Galaxies were added to the simulation \textit{post facto} using semianalytic techniques \citep[e.g.][]{delucia07,guo10a,guo11,guo13a}.  Galaxies are initially associated with individual dark matter halos, which may become subhalos of larger structures over time. 
The techniques of \citet{delucia07}, for example, expanded on the methods of \citet{croton06} and incorporated the effects of rapid star formation and gas loss in galaxy mergers, changes in galaxy properties with varying stellar mass functions and stellar populations, and attenuation due to dust.

The simulation results were saved in a collection of 64 redshift ``snapshots.''  The identified dark matter halos were then assigned galaxy properties using the semianalytic techniques discussed in the references above. The snapshots, combined with the high resolution of the simulation and the sophisticated semianalytic galaxy formation and evolution models, provide an excellent tool for studying the history of CGs over cosmic time. Throughout this paper, we adopt the \citet{delucia07} catalog of galaxies created through this semianalytic approach.

\subsection{Previous Work}

A number of previous studies have utilized the Millennium Simulation galaxy catalogs to investigate CGs. These studies have been limited to relatively low redshifts to facilitate comparison to observations, such as the well-known Hickson catalog \citep{hickson82,hickson92}.  By using a light cone in the Millennium Simulation galaxy catalogs \citep{henriques12}, it has also been possible to create ``mock'' catalogs of CGs, which can be compared to observations and are subject to analogous limitations
\citep[e.g., interloping foreground or background galaxies;][]{mcconnachie08,diaz-gimenez10,diaz-gimenez15, farhang17}. 

By comparing mock CGs in projection in the Millennium Simulation to observed Hickson compact groups (HCGs), \citet{mcconnachie08} found that $\sim$29\% of mock CGs identified from ``images'' alone are physically dense systems of three or more galaxies, and that the remaining projected CGs are the result of chance alignments.  
\citet{diaz-gimenez10} found that the fractions of 3D dense groups among 2D CGs depend on the semianalytical model (SAM) used, the consideration of galaxy blending, and the criterion used to define a dense group. Their Table 5 indicates that, for particle-based group definitions and their optimal definition of a dense group combining line-of-sight size and elongation, the fractions of CGs selected in projection that are physically dense range from 20\%\ for the \citet{delucia07} SAM to 43\%\ for the \citet{bower06} SAM. These fractions rise to over half and up to 3/4, 
respectively, for CGs that survive the velocity filter. The differences with the results from the McConnachie team arise from different definitions of what constitutes a physically dense group.

Finally, by comparing CGs observed in the Two Micron All Sky Survey (2MASS) catalog to those in a mock light cone from a Millennium Simulation {galaxy catalog}, \citet{diaz-gimenez15} found that only about a third of CGs are embedded in larger structures, i.e., the majority are truly isolated systems, yet \citet{diaz-gimenez10} find that only 11\%\ of their velocity-filtered CGs are constituted by the Friends-of-Friends group in 3D and hence are isolated.

\section{Identification of CGs in the Millennium Simulation Galaxy Catalogs}

Identifying CGs with robust criteria has presented a challenge to investigators in a number of studies, and the resulting group demographics are sensitive to these criteria (\citealt[e.g.][]{geller83, nolthenius87}; see also
Table 5 in \citealt{diaz-gimenez12}, \citealt{taverna16} and references therein). Here we invoke an algorithm with a set of tunable parameters that allows us to investigate how the demographics of a CG sample depend on individual parameters.

\subsection{Algorithm \label{algorithm}}

\begin{figure}
    \centering
    \includegraphics[width=\columnwidth]{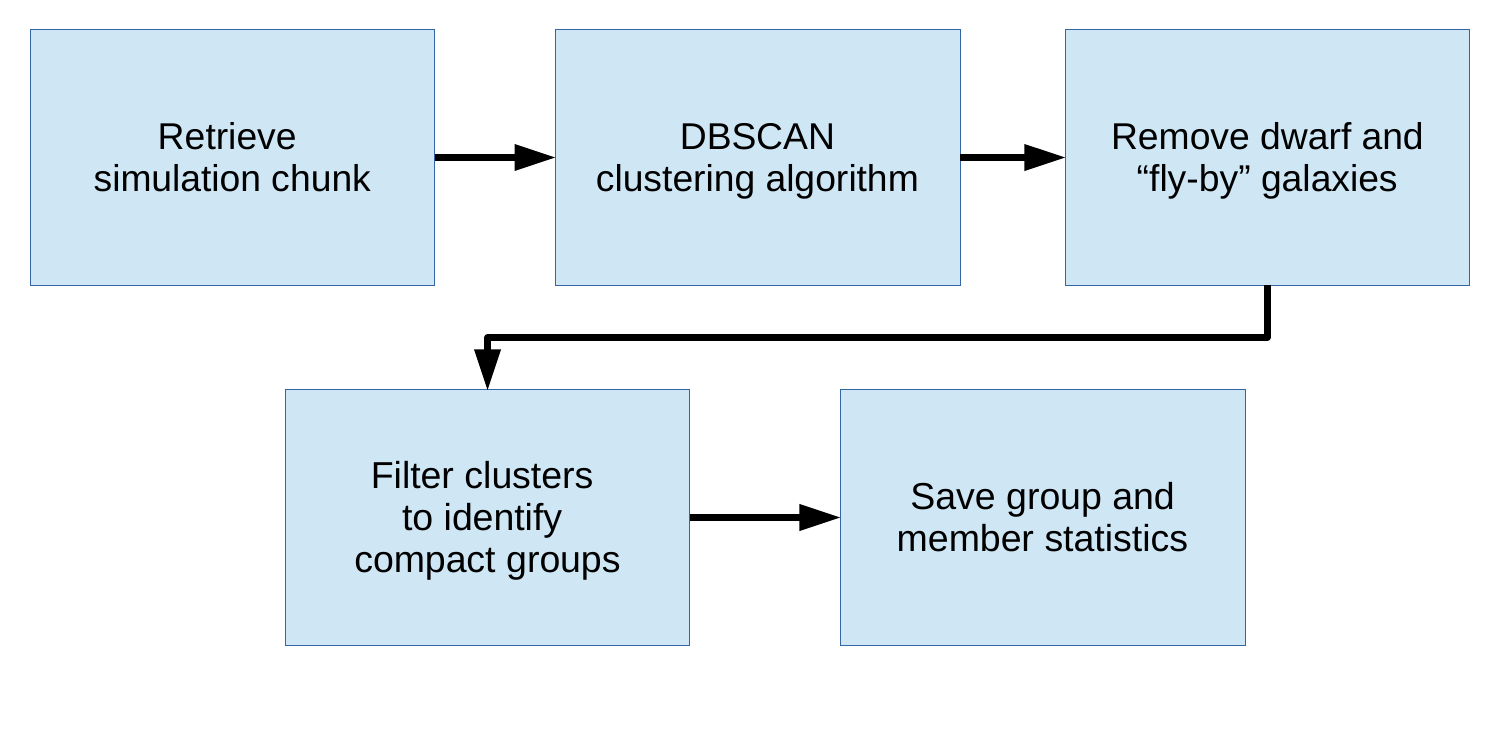}
    \caption{\footnotesize Algorithmic process developed for obtaining and analyzing Millennium Simulation galaxy catalogs.}
    \label{fig:algorithm}
\end{figure}

We developed an open-source, publicly available code\footnote{
\href{https://ascl.net/1811.010}{https://ascl.net/1811.010}} to identify and characterize CGs in Millennium Simulation galaxy catalogs. This process is subdivided into several simple algorithms with tunable parameters defining the properties of galaxy groups considered ``compact,'' where we use ``compact'' to refer to the 3D spatial separation. 
The general outline
of our methodology
(see also Figure~\ref{fig:algorithm}) is as follows:
\begin{enumerate}
\itemsep-4pt 
\item Download galaxy data from a chunk (snapshot at a given redshift)
of the Millennium Simulation, utilizing the \citet{delucia07} catalog.
\item Use a tunable clustering algorithm to identify clusters and groups of galaxies in the simulation chunk.
\item Filter out identified group members that are dwarf or ``fly-by'' galaxies.
\item Measure statistics of the discovered galaxy groups.
\item Use tunable parameters to filter the identified galaxy groups to select only ``CGs.''
\item Save the statistics of identified CGs and their galaxy members.
\end{enumerate}

These algorithms contain several tunable parameters that can be used to adjust how CGs are selected from the Millennium Simulation galaxy data. These are briefly outlined here and described in greater detail in the following sections:
\begin{enumerate}
\itemsep-4pt 

\item Neighborhood, {\sc nh}---the neighborhood parameter of the {\sc DBScan} algorithm (similar to the ``linking length'' parameter of friends-of-friends algorithms).

\item Minimum number of members, \nmin---the minimum number of nondwarf galaxy members an identified candidate group must have in order to be considered a CG.

\item Maximum shell density ratio, 
{\sc sr}---the maximum allowable ratio of the virial mass\footnote{Virial mass is defined in \citet{delucia07} as the number of gravitationally bound dark matter particles present in the subhalo.} density of galaxy subhalos in a shell surrounding a CG to the virial mass density of galaxy subhalos within that CG.

\item Galaxy mass ratio, {\sc mr}---the minimum allowable virial mass ratio of the secondary+tertiary dark matter subhalos to the primary galaxy dark matter subhalos in a group.

\item Critical velocity difference, $|\Delta v|$---the velocity difference between a galaxy and the median velocity of a CG for the galaxy to be considered a ``fly-by'' and not a member of the CG.

\item Dwarf limit---the stellar mass limit for a galaxy to be considered a dwarf galaxy and thus excluded from consideration.

\end{enumerate}

\subsection{Tunable Clustering Algorithm}

To identify clusters and groups of galaxies, we used
{\sc DBscan}, which employs an intuitive, number-density-based clustering approach  \citep{ester96}. {\sc DBscan} is very
adept at handling arbitrary shapes, as it does not depend on any density-smoothing processes. 
In clustering terminology, each galaxy constitutes a ``node.''
The algorithm works by searching the defined 
radial
neighborhood of each node and counting the number of neighbors that are within this radius. If the neighborhood contains more than, or equal to, the
minimum number of members, then it is flagged.
This is repeated for each node/member until 
all nodes within the neighborhood have been searched,
thus classifying the resulting flagged collection of nodes as a ``cluster'' \citep{birant06}, which in our case is an identified CG. 
The group's center is defined as the median of the positions of the identified galaxy members. The group radius is then the greatest distance from this center to any of the member galaxies.

Thus, {\sc DBscan} requires two initial parameters, {\it neighborhood} and {\it minimum number of members}. {\it Neighborhood} is the radius starting on each identified node and used to search for adjacent nodes. {\it Minimum number of members} specifies the required minimum number (density) of galaxies that are in the neighborhood of an identified node (a ``central'' galaxy), for the system to be considered a CG.

One caveat about the {\sc DBscan} algorithm is that is only deterministic under certain conditions. The minimum number of members must be less than or equal to three. If this condition is met, which it is in this study, then the algorithm is fully deterministic.

\subsection{Properties of HCGs}

We selected our CGs in 3D space from the SAM model outputs in a similar manner to how Hickson selected his CGs in redshift space. The selection criteria Hickson used were based on compactness, relative luminosity, minimum number of members, and isolation. The particular parameters were chosen to identify dense systems of multiple galaxies while excluding substructure within galaxy clusters.

Using imaging alone, \citet{hickson82} 
defined his CGs with a membership of at least four galaxies
(although subsequent studies relaxed this restriction to three members; \citealt{barton96}). 
The requirement for compactness was that the surface brightness averaged over the smallest circle enclosing the group galaxy centers be brighter than 26 mag arcsec$^{-2}$. The isolation criterion was that there should be no galaxies brighter than the faintest galaxy within 3 times the angular radius of the group. This aimed to exclude CGs that are associated with galaxy clusters or other regions whose ``external'' galaxies may strongly influence group galaxies. As these restrictions are subject to projection effects, the presence of foreground or background galaxies can influence the initial identification of CGs \citep[by falsely including or excluding them;][] {mamon86, mcconnachie08, brasseur09, diaz-gimenez10}. 

In addition, the spectroscopic study of \citet{hickson92} introduced the further criterion that an ``accordant'' member galaxy should be within 1000~km~s$^{-1}$ of the median group velocity. This criterion was meant to remove distant foreground and/or background galaxies that appear in projection along the line of sight.

\subsection{Selection Criteria}

Our selection criteria were motivated by those used for Hickson's original catalog, modified to take full advantage of the 3D information in the Millennium Simulation galaxy catalogs produced by the SAM of \citet{delucia07}.

We describe subsequently in this section our choices for the input parameter values to {\sc DBScan} based on the known properties of HCGs that have been determined empirically since their identification.
The {\sc nh} parameter is the search radius around a galaxy,
effectively determining the degree of group compactness, and is the initial parameter used by the {\sc DBscan} algorithm to identify clustered galaxies. {\sc sr} is a group isolation criterion, while {\sc mr} is a criterion characterizing the dominance of the most massive galaxy in a group among the top-ranked group galaxies.

\subsubsection{Neighborhood Parameter: {\sc nh} \label{sec:neighborhood}}
The groups that were identified as HCGs have a median 2D galaxy--galaxy separation of 39~$h^{-1}$ kpc \citep{hickson92}. This separation between galaxies is related to the compactness selection criterion, since CGs have to have a surface brightness of at least 26 mag arcsec$^{-2}$ \citep{hickson82}. Note that the galaxy separation is projected from 3D space down to 2D space and therefore underestimates the 
intrinsic separation of galaxies in a group. To find the 3D correction factor, we performed a Monte Carlo simulation in which we constructed $10^6$ realizations of galaxies, randomly positioned in 3D space. The median distance between galaxies was measured in 2D space for each realization. We find that the median 3D distance between galaxies is roughly $2\pi/5$ larger than the projected 2D distance. Based on the median projected distance and the multiplier, we estimate the median separation between HCG galaxies in 3D space to be about $50 \ h^{-1}$kpc.  This directly relates to the neighborhood parameter ({\sc nh}) in the {\sc DBscan} algorithm. We 
adopt {\sc nh}~$ = 25, 50, $ and 75~$h^{-1}$ kpc to bracket the empirical value.

\subsubsection{Minimum Number of Members: \nmin } \label{sec:nmin}
There is no consensus in the literature regarding the minimum number of member galaxies for a candidate CG. The original \citet{hickson82} criterion was \nmin~$=4$, and this was adopted by a number of works \citep[e.g.][]{mcconnachie08,diaz-gimenez15}. On the other hand, several studies have included triplets \citep[\nmin~$=3$; e.g.][]{barton96,tzanavaris10,lenkic16}. We choose to include triplets as a more general lower limit for the number of CG member galaxies. We do not vary \nmin, as this would be a separate full study, the computational demands of which are well beyond this paper. For reference, we note that among our selected candidate CGs, the fractions of systems that are triplets are 61.5\%\ at $z\sim2$ and 68\%\ at $z=0$.

\subsubsection{Limit on Mass in Surrounding Shell: {\sc sr}}

In order to select against groups that might be associated with larger clusters, we invoke a maximum stellar mass--density requirement for a shell surrounding a CG. Specifically, the ratio of the stellar mass density within the surrounding shell to the stellar mass density of the CG, {\sc sr}~$\equiv {\rho_{\rm shell}}/{\rho_{\rm grp}}$, must be less than a threshold value. In other words, we compare the stellar mass density (as determined from the combined virial masses of the constituent galaxy subhalos) in a shell around an identified group to the stellar mass density of the group's identified member galaxies. We choose threshold values of log({\sc sr})~$ = -3$, $-4$, and $-5$.

Initially, we adopted a shell radial width that was a multiple of the group radius, aiming to mimic the \citet{hickson82} isolation criterion, since that was dependent on a group's angular radius. However, this led to many selected groups that were in fact located within galaxy clusters. The likely explanation for this is that some group radii were so small that multiplying them by a small factor did not adequately sample the local environment.  Instead, we opted 
to use a fixed radial width
based on (1) the distance beyond which quenching is not observed in dwarf galaxies \citep[$>$1.5~Mpc;][]{geha12} and (2) the current distance between the Milky Way and the Andromeda galaxy ($\sim 0.8$~Mpc), which roughly defines the ``core'' of the Local Group.  Based on these two empirical benchmarks, we 
adopt an intermediate shell radial width of 1~\hm\ Mpc extending beyond the identified group radius. 

Because we use the mass density at a fixed radius, we also exclude any group within 1~\hm\ Mpc from the edges of the box in each chunk, due to the fact that some volume of the sphere outside of the cube is inherently empty and skews the ratio toward a less dense environment. 

\vspace{.5cm}
\subsubsection{Mass Ratio of Group Members: {\sc mr}\label{mass_restriction}}

In order to select against groups that consist of a single massive galaxy with small satellites, \citet{hickson82} required that constituent galaxies have magnitudes within 3 mag of the brightest member. This requirement also helps to mitigate against galaxies that only appear to be members of the group due to projection effects, but are in fact at very different distances.  In the Millennium Simulation galaxy catalogs all spatial relationships between galaxies are known, ensuring that projection effects do not contaminate the sample. Another advantage of the Millennium Simulation is that it provides 
masses of the galaxy subhalos from galaxy catalogs that allowed us to directly compare the masses instead of relying on observables such as apparent magnitude \citep{delucia07}. As a result, we are much less sensitive to the specific prescriptions for star formation within the semianalytical modeling, and our results should be generalizable to other galaxy catalogs based on the Millennium Simulation. 

\begin{figure}[ht]
 \vspace{-5.7cm}
   \hspace{-10pt}
    \includegraphics[width=1.1\columnwidth]{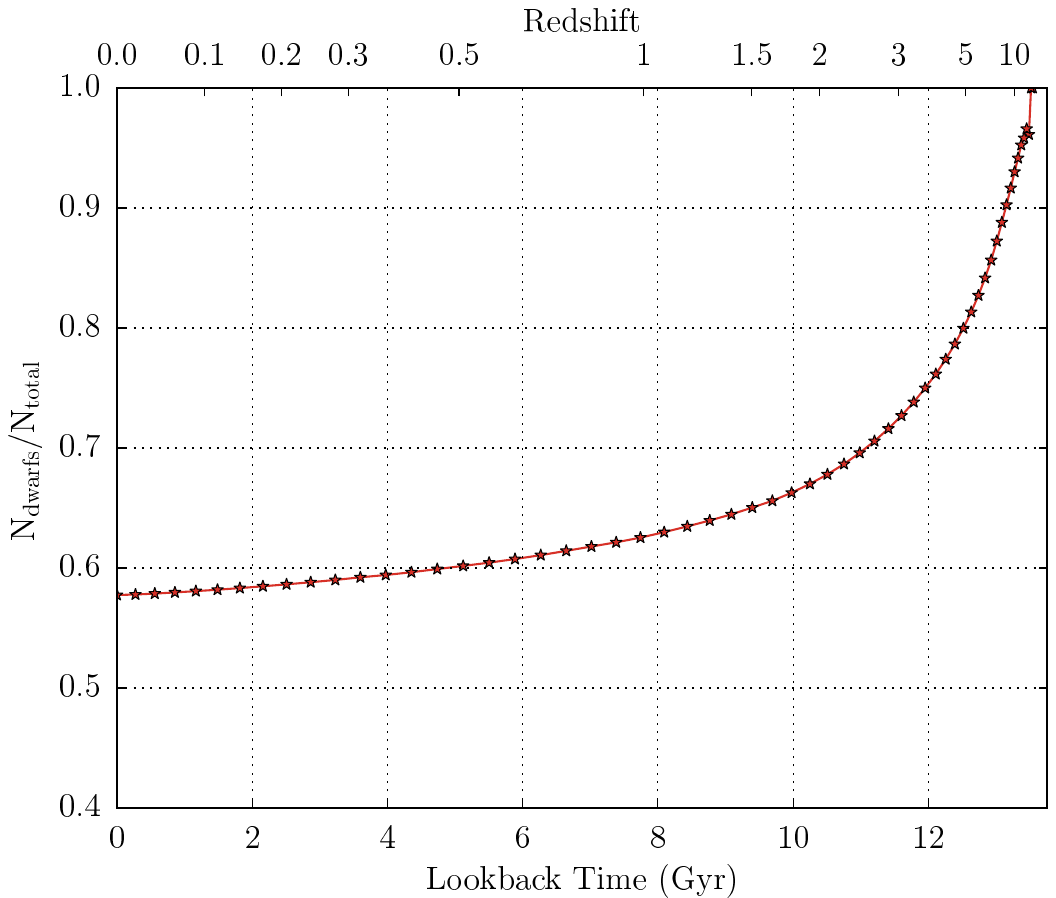}
    \caption{\footnotesize \hspace{-10pt}Evolution of the fraction of ``dwarf'' galaxies (stellar mass $<$5$\times 10^{8} \ h^{-1} \ M_{\odot}$) identified in the \citet{delucia07} semianalytical model of galaxy formation run on the Millennium Simulation.  These systems are excluded from further analysis for the reasons discussed in \scr~\ref{dwarf_exclusion}. The exclusion of dwarf galaxies results in a lower limit on the number of identified CGs that becomes increasingly more significant at higher redshift.  For $z<0.5$, the fraction of excluded galaxies based on the mass criterion is $\lesssim$60\%.  The fraction rises rapidly for $z>1$. See \scr~\ref{dwarf_exclusion} for further discussion.}
    \label{fig:dwarf_fraction}
\end{figure}

We make the assumption that dominance in galaxy luminosity corresponds to dominance in galaxy and subhalo mass. 
In order to select against groups that consist of a single dominant galaxy with minor satellites, we required a minimum, threshold value for the mass ratio of the combined mass of the second and third most massive galaxies to that of the most massive galaxy, defined as 
{\sc mr}~$\equiv {(M_{\rm secondary} + M_{\rm tertiary}) / M_{\rm primary}}$.
While this requirement still allows for ``minor'' galaxy companions, it excludes systems similar to the Milky Way and its satellites. The advantage of this method, 
as opposed to a hard cutoff in galaxy subhalo mass,
is that it allows less luminous galaxies to be members of a CG as long as they are sufficiently massive. We choose minimum threshold values of {\sc mr}~$= 0.20$, {\sc mr}~$=0.10$, and {\sc mr}~$=0.05$; a three-``magnitude'' mass ratio would correspond to {\sc mr}~$= 0.06$.

\subsubsection{Relative Velocity Restriction}
To avoid identifying galaxies as members of a CG that are not subject to a long-term dynamical interaction, we introduce a velocity filter. 
This filter eliminates any galaxies that have $\mid\Delta v\mid\ > 1000$~km~s$^{-1}$ from the median group velocity and is the velocity filter that was used by \citet{hickson92}.  
We find that this only excludes $1.6\%$ of potential CGs.
We note that for observational studies, galaxies with peculiar velocities could appear to have accordant velocities but still potentially not be physically associated with a group. Given that all spatial relationships are known for the galaxies in this work, group members are not subject to this caveat.  

\subsubsection{Exclusion of Dwarf Galaxies \label{dwarf_exclusion}}
In our analysis we impose a galaxy mass threshold to exclude galaxies with stellar masses that may be too low. Specifically, only galaxies that have masses greater than $5\times 10^8 h^{-1} M_\odot$ are included. This corresponds to a median dark matter halo mass of $3.4\times10^{10} M_\odot$ or $\sim 40$ dark matter particles.
 Any ``galaxies'' with zero stellar mass are also removed from the analysis, as they are artifacts of the \citet{delucia07} SAM.
As shown in Figure~\ref{fig:dwarf_fraction}, even for redshifts of $z\lesssim0.5$, this results in the exclusion of $\sim 60$\% of the mass systems identified in the Millennium Simulation galaxy catalog. At earlier cosmic times, the relative fraction of such ``dwarf'' galaxies that are excluded owing to this mass cutoff becomes increasingly important, reaching more than 80\%\ at $z \gtrsim 5$.

\subsubsection{Significance of Selection Criteria}
In order to both mitigate the effect of a set of rigid selection criteria and also investigate how CG demographics may depend on these criteria, we vary {\sc nh, sr}, and {\sc mr}
over a range of values about a ``default'' set (see Table~\ref{parameter_sets}). We first test each non-default value by keeping all other criteria at their default values and then also use combinations of the most ``lenient'' and most ``restrictive'' parameter values, in order to constrain the most extreme CG populations.
In the case of {\sc nh}, lower values reduce the size of the search area for neighboring galaxies, so that more CGs will be selected, and conversely for higher values. For very large values, galaxy clusters will start to be included. Our default value is $50 \, h^{-1} \, {\rm kpc}$ (see Section~\ref{sec:neighborhood}), and we also use the values 25 and $75 \, h^{-1} \, {\rm kpc}$.

Our default value for log({\sc sr}) is $-4$, and we also use $-3$ and $-5$. For {\sc mr} the default value is 0.10, and we also use 0.05 and 0.20.

\section{Processing Algorithm}

\subsection{Optimization Strategy}
To be able to analyze the full simulation in a time-efficient manner, we developed optimization strategies. 
The resulting time requirements are extreme: the full simulation catalog has over 26 million galaxies in its most populous snapshot. In order to make the algorithm more computationally efficient, we divide a given snapshot into roughly equivalent overlapping boxes. Each box has a side length of 102~\hm\ Mpc, corresponding to about $1/125$ of the total simulation volume. Only the central 100~\hm\ Mpc were sampled in each box, since, when conducting the density analysis, a buffer zone of 1~\hm\ Mpc is needed at the edge. Without this buffer, the groups on the edge of the box would contain less volume in the spherical density analysis than required to determine a group's isolation. This process of division allowed us to run the algorithm in an extensively parallelized fashion on as many cores as desired.

\vspace{-0.cm}
\begin{figure*}[ht]
  \centering
    \includegraphics[width=0.45\textwidth]{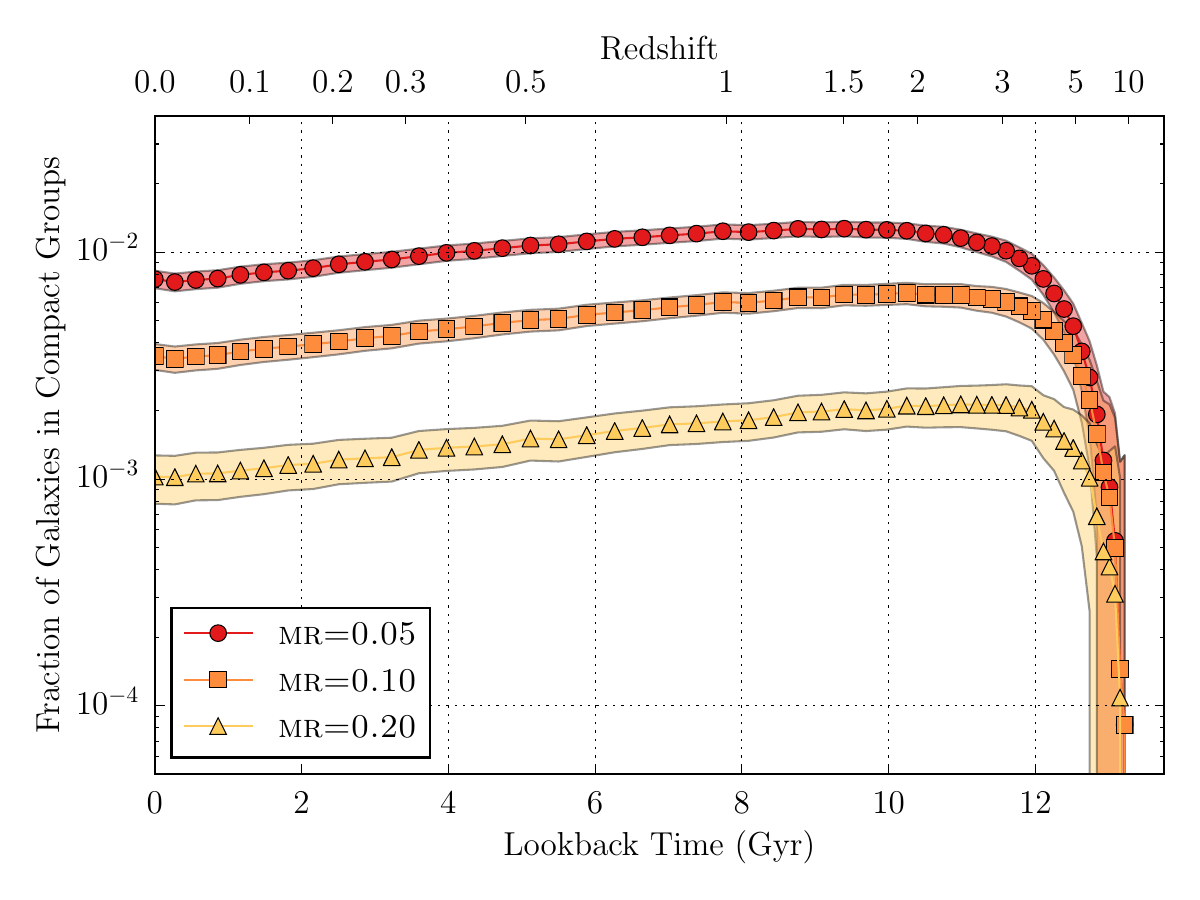}
    \includegraphics[width=0.45\textwidth]{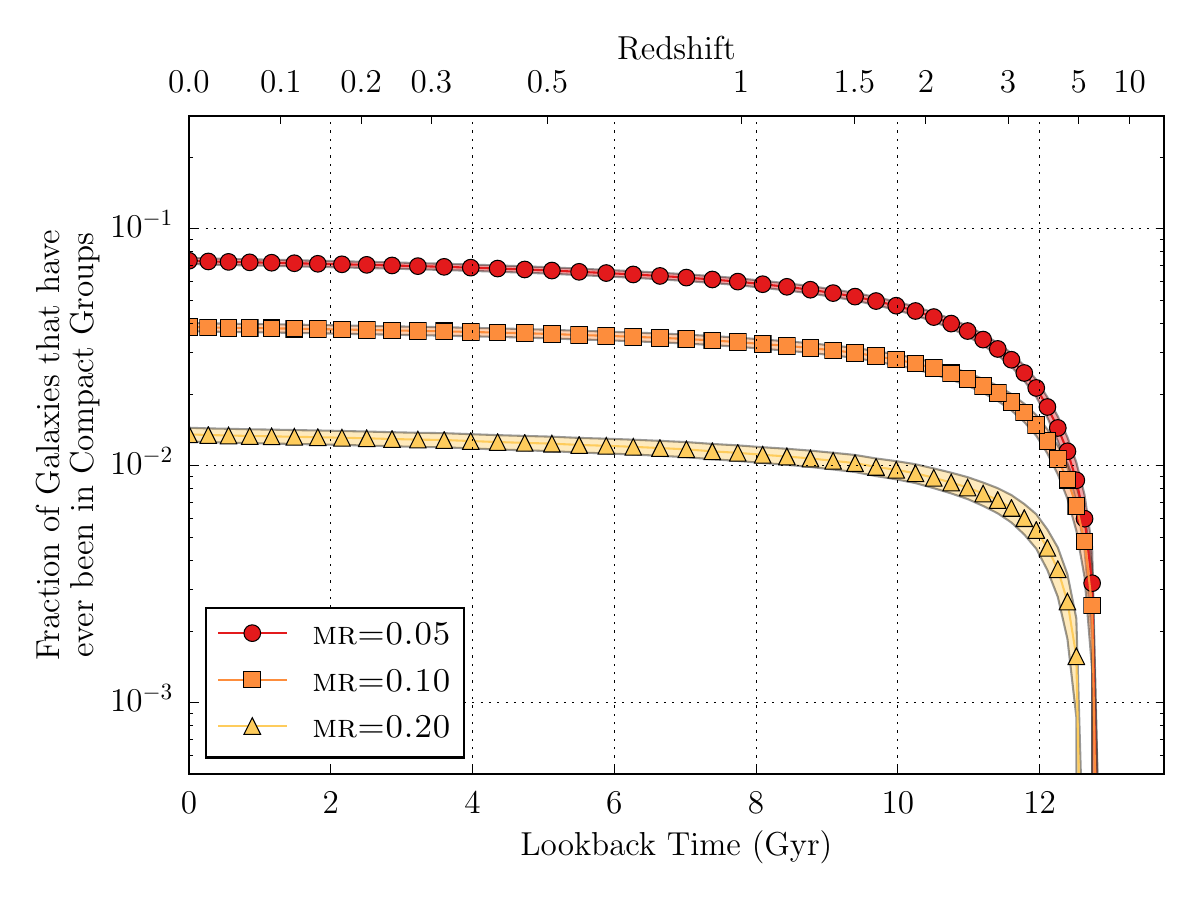}
   \caption{\footnotesize Fraction of galaxies (stellar mass greater than $5\times 10^8 h^{-1} M_{\odot}$) living in CGs (left) or that ever lived in one (right). Each analysis uses a neighborhood parameter {\sc nh} = 50~\hm\ kpc and a maximum shell density ratio \(\log_{10}(\text{\sc sr}) = -4\). These curves are the results of varying the mass ratio (\(\text{\sc mr} = 0.05\), red circles; \(\text{\sc mr} = 0.10\), orange squares; \(\text{\sc mr} = 0.20\), yellow triangles). The width of the shaded region is 50 times the Poisson uncertainty. In all cases, the fractions of galaxies in CGs peak at $z\sim$1.0--2.4, and drop off sharply at $z\sim5$ because of our dwarf galaxy cutoff mass.  As expected, the most restrictive value of {\sc mr}~$=0.20$ (with top-ranked galaxies of more comparable masses) results in fewer CGs at all redshifts. }
    \label{fig:secondtwo}
\end{figure*}

\renewcommand{\arraystretch}{0.89}
\begin{deluxetable*}{lcccccccccc}[t]  
\tablecolumns{11}
\setlength{\tabcolsep}{7pt}
\tabletypesize{\footnotesize}
\tablecaption{{\footnotesize Parameter Grid and Results}
\label{parameter_sets}}
\tablehead{   
  \colhead{\hspace{-10pt}Parameter} &
  \colhead{\sc nh} &
  \colhead{log(\sc sr)} &
  \colhead{\sc mr} &
  \colhead{Median Galaxy} &
  \colhead{Median Group} &
  \colhead{Median $\sigma_{v,\rm 3D}$} &
  \colhead{Max.} &
  \colhead{$z$} &
  \colhead{$N_{\rm groups}$} &
  \colhead{$n_{\rm groups}$}
  \\
  \colhead{\hspace{-32pt}Set} &
  \colhead{} &
  \colhead{} &
  \colhead{} &
  \colhead{Separation } &
  \colhead{Radius } &
  \colhead{(3D)} &
  \colhead{Pop.} &
  \colhead{of Peak} &
  \colhead{at~$z=0$} &
  \colhead{at~$z=0$}
  \\
  \colhead{} &
  \colhead{(\hm\ kpc)} &
  \colhead{} &
  \colhead{} &
  \colhead{(\hm\ kpc)} &
  \colhead{(\hm\ kpc)} &
  \colhead{(km s$^{-1}$)} &
  \colhead{} &
  \colhead{} &
  \colhead{} &
  \colhead{($10^{-5}\, h^3 \, {\rm Mpc}^{-3}$)}
\\  
  \midrule[0.8pt]
  \colhead{\hspace{-35pt}(1)} &
  \colhead{(2)} &
  \colhead{(3)} &
  \colhead{(4)} &
  \colhead{(5)} &
  \colhead{(6)} &
  \colhead{(7)} &
  \colhead{(8)} &
  \colhead{(9)} &
  \colhead{(10)}&
  \colhead{(11)}
}
\startdata
\hspace{-5.6pt}A   & 25  & $-4$  &   0.10  &  19   &  17 & 99  &  0.18\%  &   1.0 & 3854    & 3.1      \\
\hspace{-5.6pt}B   & 75  & $-4$  &   0.10  &  47   &  45 & 95  &  1.03\%  &   1.9 & 16708   & 13.5     \\
\hspace{-5.6pt}C   & 50  & $-3$  &   0.10  &  37   &  34 & 102 &  0.83\%  &   1.9 & 13940   & 11.3     \\
\hspace{-5.6pt}D   & 50  & $-5$  &   0.10  &  29   &  26 & 97  &  0.29\%  &   1.5 & 5973    & 4.8      \\
\hspace{-5.6pt}E   & 50  & $-4$  &   0.05  &  37   &  34 & 118 &  1.27\%  &   1.5 & 24134   & 19.5     \\
\hspace{-5.6pt}F   & 50  & $-4$  &   0.20  &  35   &  31 & 83  &  0.21\%  &   2.4 & 3355    & 2.7      \\
\midrule[0.8pt]
\hspace{-5.6pt}Default      & 50  & $-4$  &   0.10  &  37  &  33 & 99  &  0.66\%  &   1.9 & 11222   & 9.1   \\
\hspace{-5.6pt}Restrictive  & 25  & $-5$  &   0.20  &  18  &  17 & 82  &  0.05\%  &   1.4 & 1032    & 0.8   \\
\hspace{-5.6pt}Lenient      & 75  & $-3$  &   0.05  &  57  &  53 & 119 &  3.13\%  &   1.5 & 53288   & 43.1  \\
\midrule[0.8pt]
\hspace{-5.6pt}HCGs&\ldots&\ldots&\ldots   &39 &  35 & 331 &\ldots   &\ldots & \ldots  & 9.5      \\
\enddata
\hspace{10pt}\tablecomments{\footnotesize Column (1): label of the observational or computational parameter set, characterized by the parameter values that were varied, as specified in Columns (2), (3), and (4). Columns (2)--(4): {\sc DBscan} algorithm neighborhood radius, maximum mass-density ratio of galaxies in a shell surrounding a candidate group to the candidate group's galaxies, and minimum mass ratio of the second and third galaxies combined to the most massive galaxy in a candidate group, respectively. The remaining columns show results for HCGs or CGs identified in the \citet{delucia07} catalogs produced from the Millennium Simulation, following the adoption of the selection criteria in Columns (2)--(4). Column (5): median galaxy--galaxy separation in a group. Column (6): median group radius.
The group radius is defined relative to the group center, taken to be the
median of the positions of the identified galaxy members. The group radius
is simply the greatest distance from this center to any of the member galaxies.
Column (7): median 3D galaxy velocity dispersion,  
$\sigma_{v,\rm 3D} \equiv \sqrt{\sigma_{v,x}^2 + \sigma_{v,y}^2 + \sigma_{v,z}^2}$. 
Column (8): maximum fraction of galaxies that were in a CG at the redshift given in Column (9). Column (9): redshift at which the number of CGs was at its maximum. Column (10): number of CGs at the present time. Column (11): volume number density of CG galaxies at the present time.
}
\vspace{-0.5cm}
\end{deluxetable*}
\renewcommand{\arraystretch}{1}

\subsection{Tracking Galaxies through the Simulation}
To track the galaxies in any snapshot that are, or have ever been, in CGs, each galaxy had to be monitored throughout the simulation. The Millennium Simulation employs a standard naming convention for galaxies and also provides the descendant ID. Once the groups are identified in the main algorithm, they can then be traced through the simulation forward or backward in time to determine the 
fraction of galaxies living in CGs at any given time step.

 Starting from the beginning of the simulation (cosmic time~$=0$), galaxies and their descendants are placed into a list, which is then searched for repeated galaxies (e.g., when a galaxy merges with another one). This process is repeated for every galaxy in a given treeID. The latter is a grouping assigned by \citet{delucia07} to categorize related groups of galaxies, so that all galaxies in a given CG will have the same treeID. 
 By dividing the data set according to treeID, further extensive parallelization was achieved.

 \begin{figure*}[ht]
   \vspace{-0.2cm}
    \centering
    \includegraphics[width=0.45\textwidth]{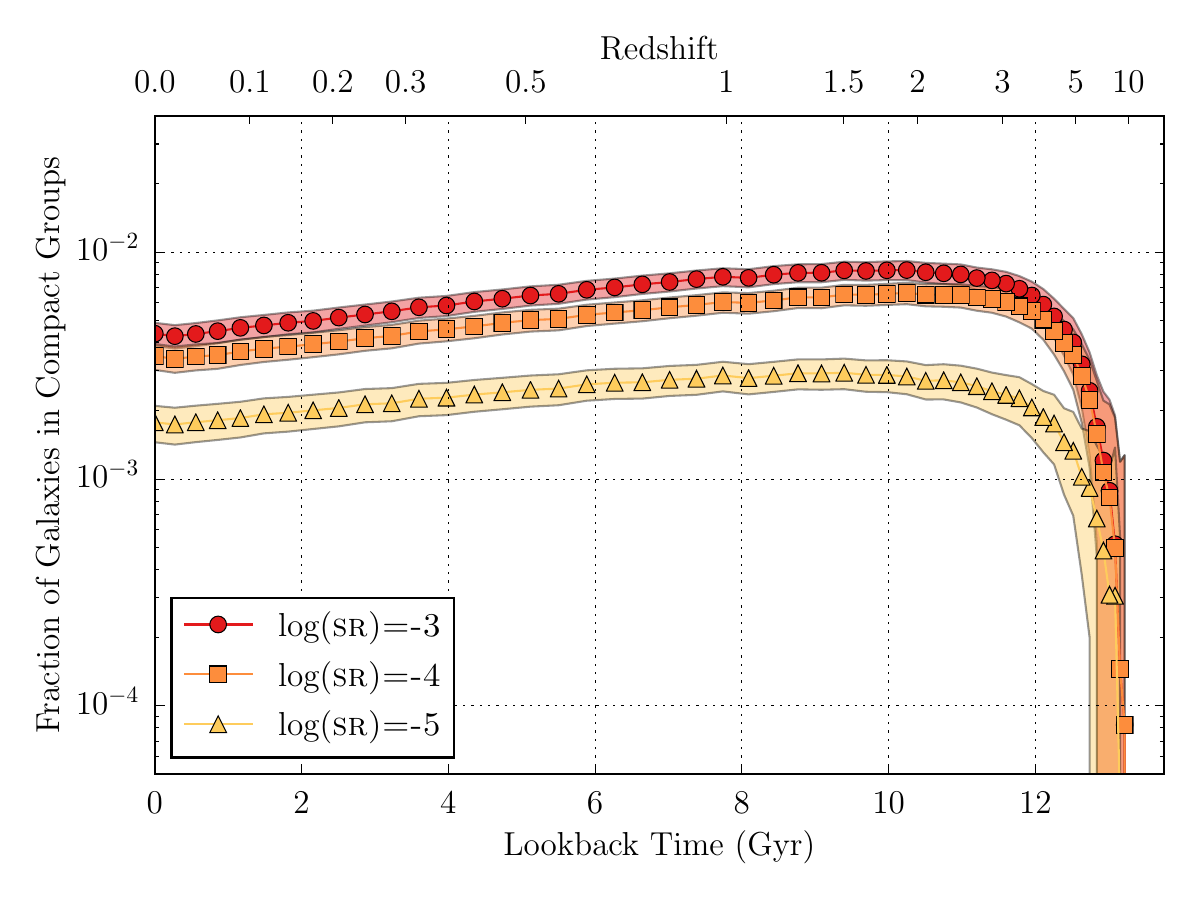}
    \includegraphics[width=0.45\textwidth]{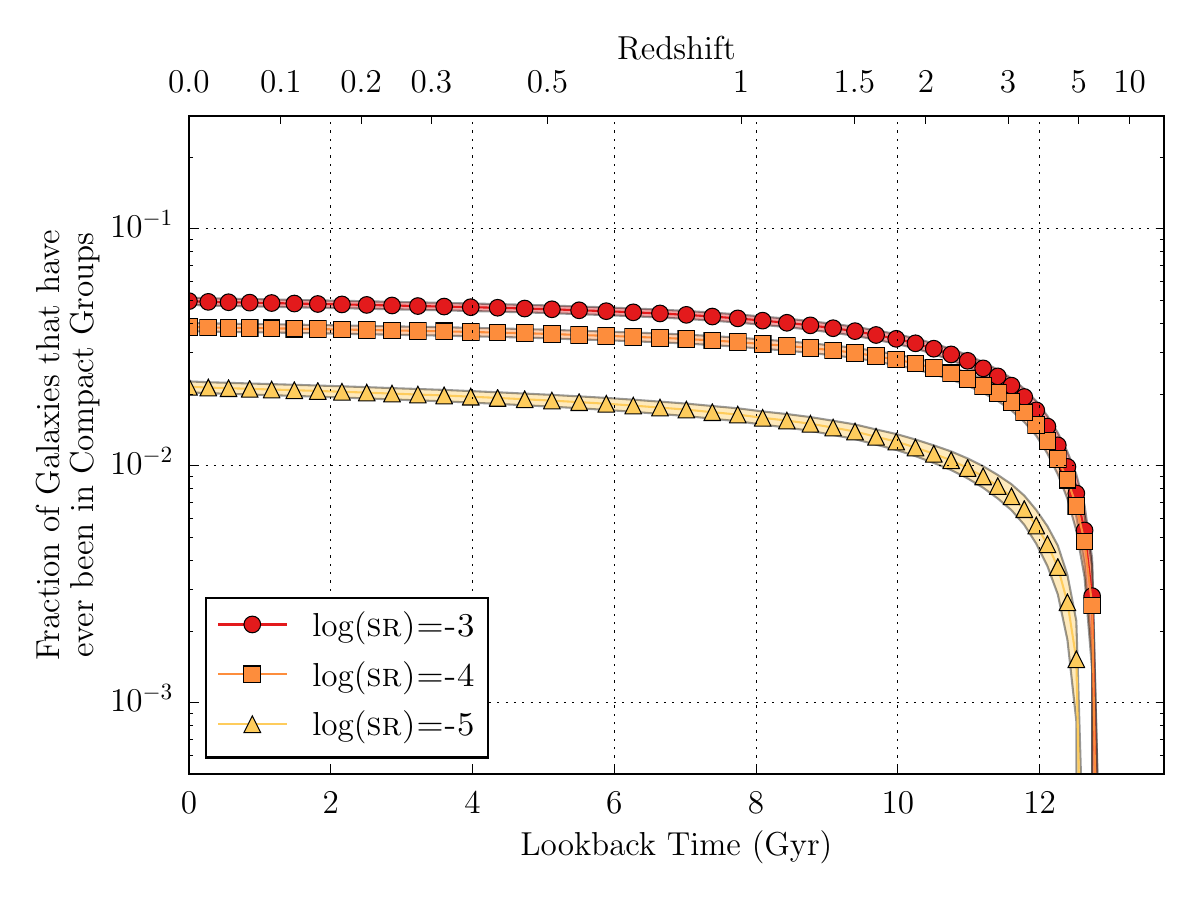}
    \caption{\footnotesize Same as Figure~\ref{fig:secondtwo} but varying the maximum shell density ratio. Each analysis uses a neighborhood parameter \(\text{\sc nh} = 50\, \) \hm\ kpc and a minimum mass ratio \(\text{\sc mr} = 0.10\). These curves are the results of varying the maximum shell density ratio (\(\log(\text{\sc sr}) = -3\), red circles; \(\log(\text{\sc sr}) = -4\), orange squares; \(\log(\text{\sc sr}) = -5\), yellow triangles). The width of the shaded region is 50 times the Poisson uncertainty. Though the overall shapes of the curves are similar to those in Figure \ref{fig:secondtwo}, varying the values of {\sc sr} by two orders of magnitude did not produce as strong of an effect as varying {\sc mr} by a factor of 4.   A restrictive {\sc sr} value of $-5$ (indicating very isolated systems) led to the fewest CG galaxies at all redshifts.    }
    \label{fig:annularmass}
\end{figure*}

\section{Results}
In this section, we present the results of multiple trials using different parameter sets and describe how the adopted selection criteria affect CG properties and demographics. 
Table~\ref{parameter_sets} presents the grid of parameter sets and resulting group properties, as well as some key observational HCG properties for comparison.

\subsection{Impact of Parameters on Group Properties}

\subsubsection{Mass Ratio of Group}
The restriction on the mass ratio of secondary+tertiary to primary galaxies ({\sc mr}) has a strong impact on the fraction of galaxies considered to be in CGs at 
$z\lesssim 5$.  
A mass ratio of {\sc mr}~$=0.05$ is the most lenient value and thus results in the largest fraction of galaxies in CGs.  Conversely, {\sc mr}~$= 0.20$ is the most restrictive and reduces the relative number of galaxies considered to be in a CG by nearly an order of magnitude for  $z< 5$ (see Figure~\ref{fig:secondtwo}).  All three values of {\sc mr} result in the fraction of galaxies in CGs peaking between $z \sim 1.5$ and 2.4, with the more restrictive values of {\sc mr} peaking at higher redshifts. The most liberal value of {\sc mr~=~0.05} (with other parameters set to their default values, set E) results in a maximum fraction of galaxy membership in CGs of $\sim$1.3\%. 

\begin{figure*}[ht]
    \centering
    \includegraphics[width=0.45\textwidth]{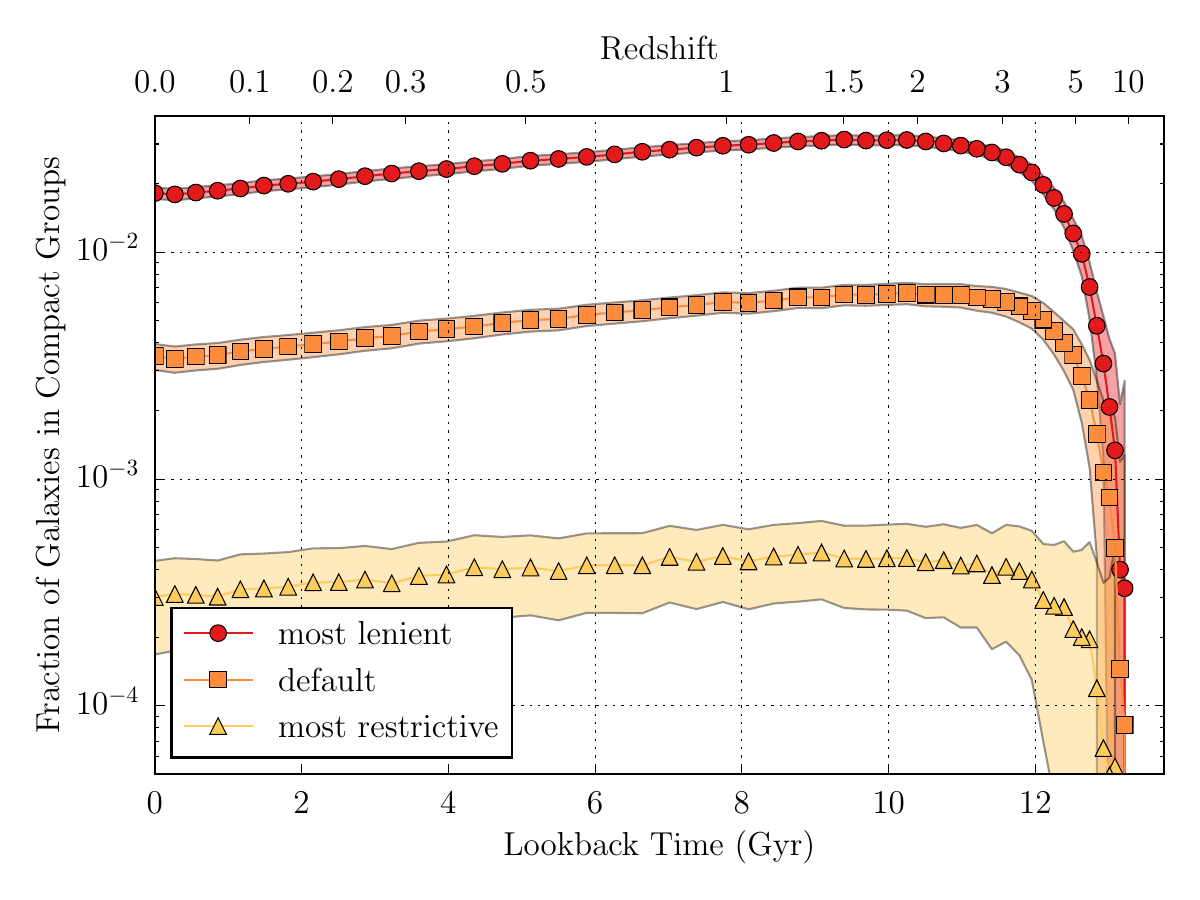}
    \includegraphics[width=0.45\textwidth]{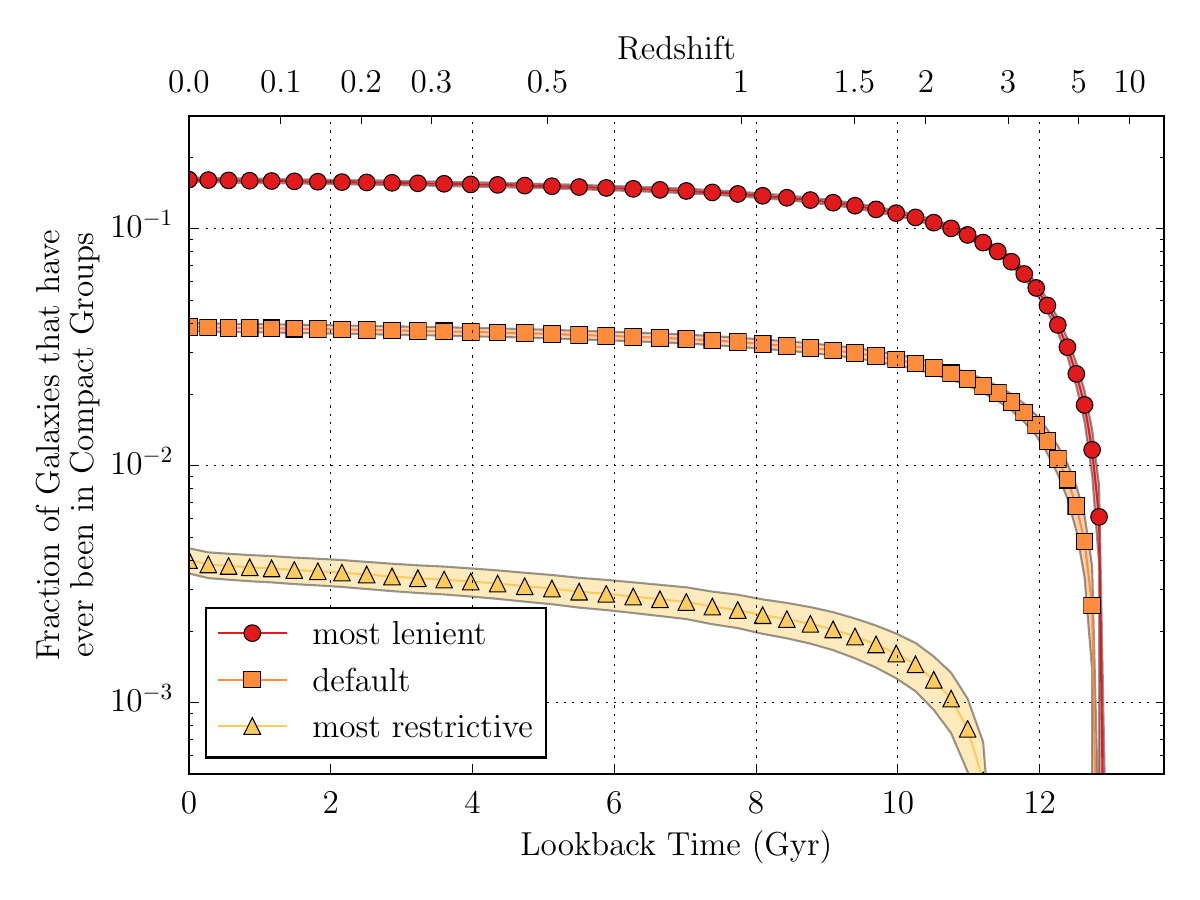}
    \caption{\footnotesize Same as Figure~\ref{fig:secondtwo} but varying all three parameters. The ``most lenient'' curve (red circles) is generated using a neighborhood parameter \(\text{\sc nh} = 75\) \hm\ kpc, a maximum shell density ratio \(\log_{10}(\text{\sc sr}) = -3\), and a mass ratio \(\text{\sc mr} = 0.05\). The ``default'' curve (orange squares) with values chosen to best approximate the HCG sample is the result using our ``default'' values of \(\text{\sc nh} = 50\) \hm\ kpc, \(\log_{10}(\text{\sc sr}) = -4\), and \(\text{\sc mr} = 0.10\). The ``most restrictive'' curve (yellow triangles) is the result using \(\text{\sc nh} = 25\) \hm\ kpc, \(\log_{10}(\text{\sc sr}) = -5\), and \(\text{\sc mr} = 0.20\). The width of the shaded region is 50 times the Poisson uncertainty. The specific input criteria for the clustering algorithm are clearly important in determining the normalization of the curves, but their shapes in both panels are similar with CG galaxy number peaking at $z\sim1.7$.}
    \label{fig:bestworst}
\end{figure*}

\begin{figure*}[ht]
    \centering
    \includegraphics[width=0.45\textwidth]{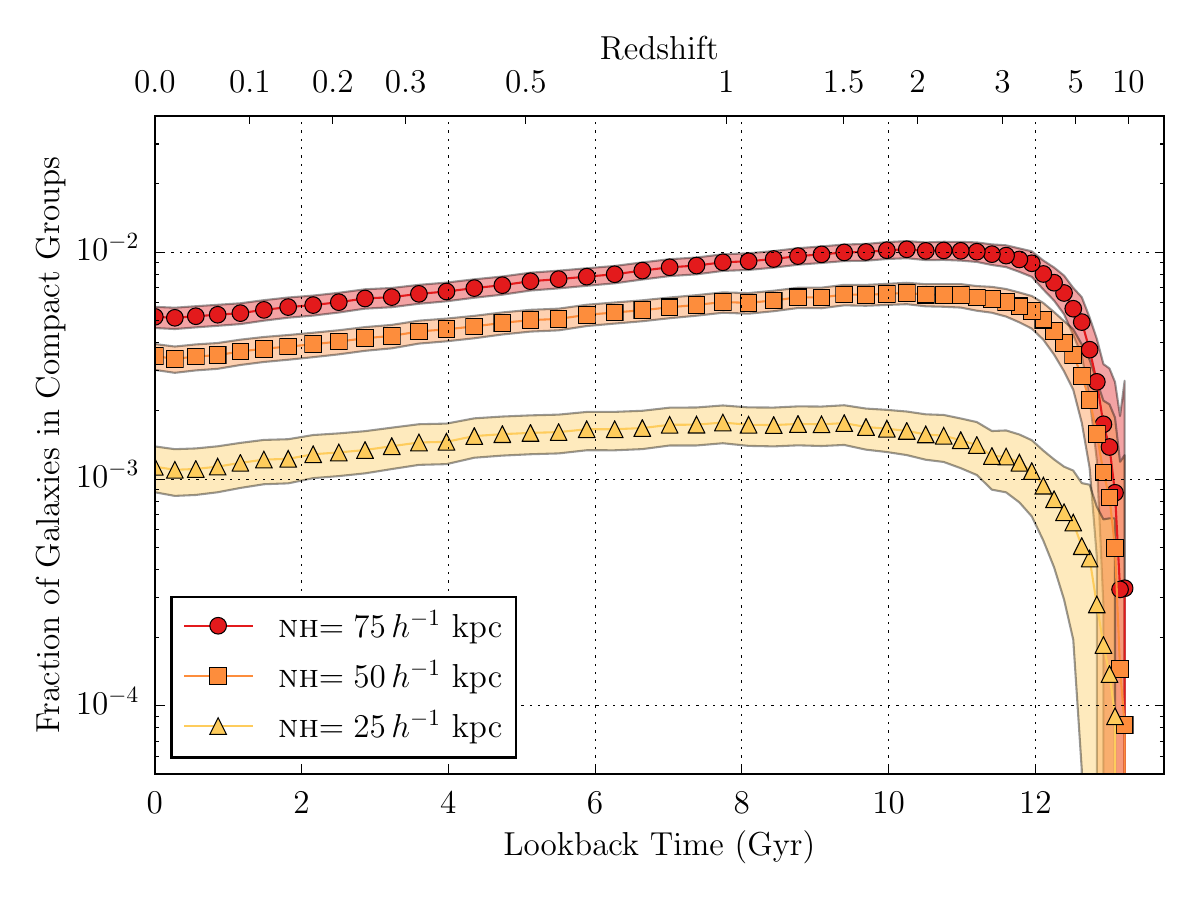}
    \includegraphics[width=0.45\textwidth]{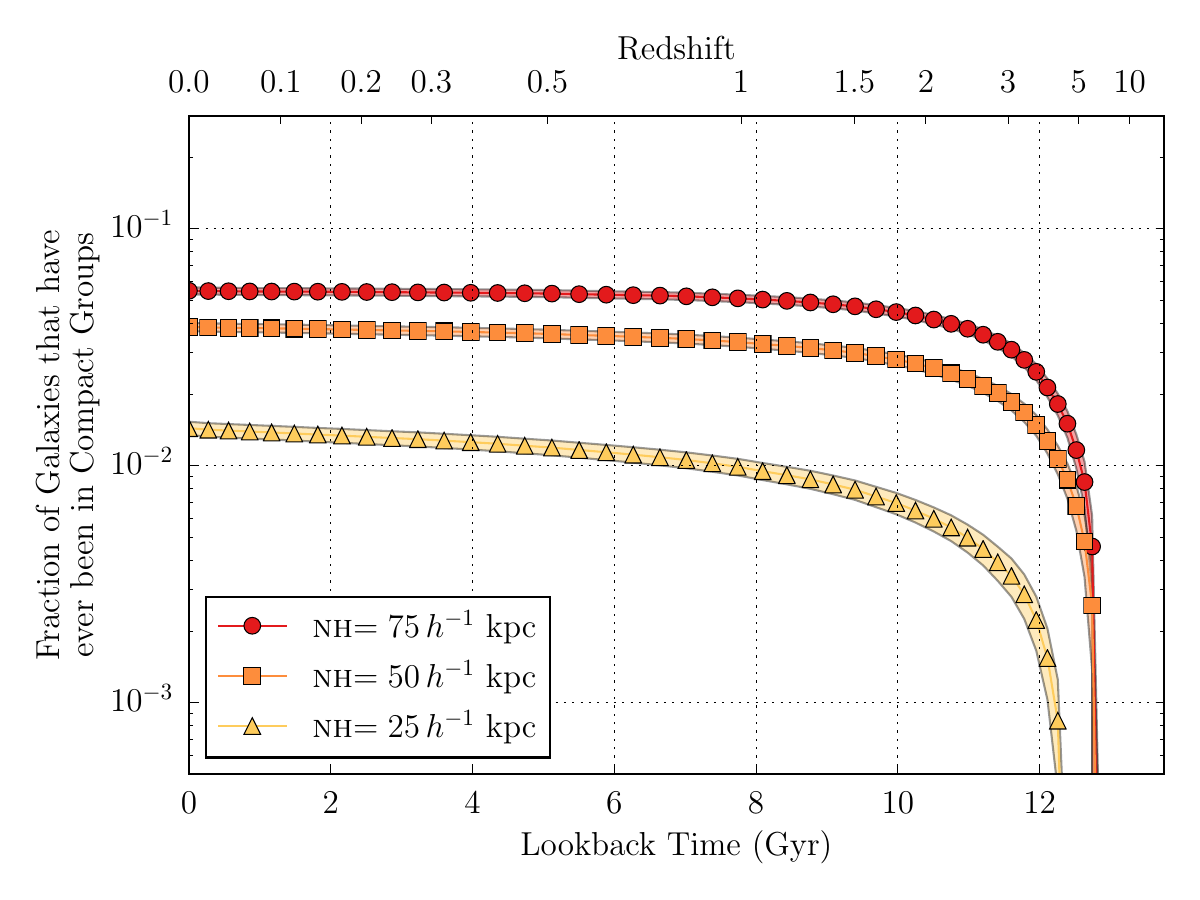}
    \caption{\footnotesize Same as Figure~\ref{fig:secondtwo} but varying the neighborhood parameter. Each analysis uses a maximum shell density ratio \(\log_{10}(\text{\sc sr}) = -4\) and a minimum mass ratio \(\text{\sc mr} = 0.10\). These curves are the results of varying the neighborhood parameter (\(\text{\sc nh} = 75\, \) \hm\ kpc, red circles; \(\text{\sc nh} = 50\) \hm\ kpc, orange squares; \(\text{\sc nh} = 75\) \hm\ kpc, yellow triangles). The width of the shaded region is 50 times the Poisson uncertainty. The most restrictive {\sc nh} value of 25 $h^{-1}$\ kpc requires identified systems to be denser.}
    \label{fig:neighborhood}
\end{figure*}

\subsubsection{Shell Density Ratio}
Varying the shell density ratio ({\sc sr}) selection criterion resulted in similar patterns to those seen with respect to changing {\sc mr}. Despite varying {\sc sr} by two orders of magnitude, the resulting impact on the fraction of galaxies in CGs is not as strong as seen from only changing {\sc mr} by a factor of four.  
However, for $z<5$ different values of {\sc sr} do
strongly influence
CG selection (Figure~\ref{fig:annularmass}).  For the most restrictive (i.e., most isolated) value of log({\sc sr})$= -5$ (model D), the relative number of galaxies in CGs never reached values greater than $\sim$0.3\%. 
The most liberal value of log({\sc sr})$ = -3$, with other parameters at their default values (Model C), results in a peak fraction of galaxies in CGs of $\sim 0.83$\% at $z = 1.9$.

\subsubsection{Neighborhood}
Variations in the neighborhood parameter ({\sc nh}) also had a strong impact on the demographics of CGs identified in the simulation.  As discussed in Section~\ref{comp_to_HCGs}, a value of {\sc nh}~$=50$~\hm\ kpc is found to produce groups that have similar median sizes to HCGs, while {\sc nh}~$=25$~\hm\ kpc and {\sc nh}~$=75$~\hm\ kpc
result in median group sizes that are smaller and larger, respectively, than observed HCGs.

\subsubsection{Most Lenient and Restrictive Parameter Sets}

In addition to varying individual selection parameters while holding the rest constant at their ``default'' values, we also test combinations of the most lenient and most restrictive parameter values in order to constrain the most extreme populations of CGs. 
The most restrictive criteria used here are log({\sc sr})~$= -5$, {\sc mr}~$ = 0.20$, and {\sc nh}~$ = 25 \, h^{-1} \,{\rm kpc}$. The most liberal values are log({\sc sr})~$ = -3$, {\sc mr}~$ = 0.05$, and {\sc nh}~$ = 75 \, h^{-1}$~kpc. As shown in Figure~\ref{fig:bestworst}, even the most lenient set of criteria only result in the relative population of galaxies in CGs peaking at $\sim 3.1$\% near a redshift of $z\sim1.5$.

\subsection{Total Fraction of Galaxies in CGs}
In addition to the relative number of galaxies that are in CGs at any given redshift, we can also determine the relative number of galaxies that are {\it currently} in or have {\it ever} been members of a CG over cosmological time.  

Here we track the galaxies that are in a CG at any given instant and continue to count them as part of the total number of galaxies as the groups evolve, even if they should no longer be considered to belong to a CG, due to mergers of constituent galaxies causing 
the group to have fewer than three members.
By tracking the CGs' descendants, we can determine the number and properties of galaxies in the present day that once were a part of a CG in their evolutionary history.

As shown in the right panel of Figure~\ref{fig:bestworst}, the maximal fraction of galaxies in the present-day universe that have ever been part of a CG exceeds 10\%. 
Unlike the rest of the parameter sets,
the most lenient and most restrictive ones
lead to a particularly large variation in
the final percentage.
The final percentages range from 0.4\% to 16.1\% 
across the full parameter range.

As can be seen in the right panels of Figures~\ref{fig:secondtwo} and \ref{fig:neighborhood}, both of the {\sc mr} and {\sc nh} criteria have a moderate effect on the final fraction. On the other hand, the right panel of Figure~\ref{fig:annularmass} shows that the {\sc sr} criteria have little effect on the final fraction of galaxies that have ever been in CGs.

\begin{figure}[ht]
\vspace{-1cm}
\hspace{-0.9cm}
    \includegraphics[width=0.60\textwidth]{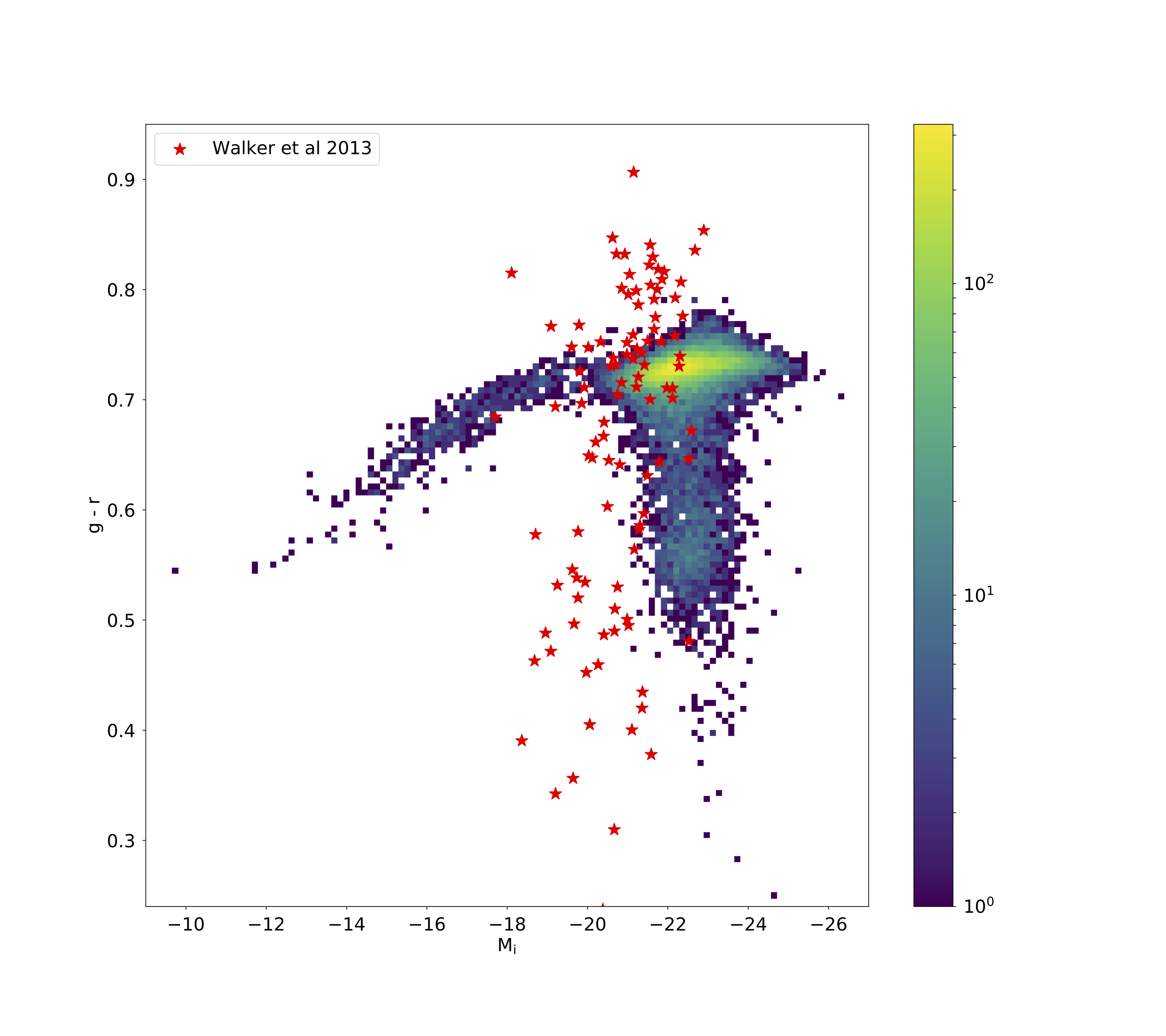}
    \vspace{-1.2cm}
    \caption{Color--magnitude diagram of Millennium catalog galaxies at $z=0$ that were in CGs at $z\sim2$. \citet{walker13} CG galaxies are overplotted as red stars.}
    \label{fig:grplot}
\end{figure}

Finally, we trace the evolution of $z=2$ CGs down to $z=0$. We find that the overwhelming majority (16,071 out of 16,797 CGs, or 96\%) have merged into a single galaxy by $z=0$.  The distribution of these galaxies at $z=0$ in color-magnitude space is shown in Figure~\ref{fig:grplot}. Taken at face value, this color-magnitude distribution is notably different from the general Sloan Digital Sky Survey (SDSS) sample presented by \citet{blanton05}; the CG descendants show a clump of red and luminous galaxies and two distinct plumes that are not seen in the general SDSS population.  However, we caution that there is no guarantee that the semianalytic model used will in general reproduce the colors of the SDSS population. Perhaps not surprisingly, this distribution is more similar to that observed for galaxies in CGs \citep{walker13}, but there remain clear differences -- including the plume of CG descendants across a range in $g-r$ color at an $I$-band magnitude of $\sim -23$. However, the caveat that applies to SDSS colors still applies here. Thus, these results tentatively and qualitatively suggest that the products of CG evolution may have properties statistically distinct from the general galaxy population, which warrants an in-depth follow-up study. 

Interestingly, we also identify a small minority of CGs (49 out of the 726 that have {\it not} merged by $z=0$) that have at least one galaxy separated by 250 kpc, or more, from other group members. Such member galaxies are reminiscent of NGC 7320C, located in the northeast quadrant of Stephan's Quintet (HCG 92) and likely to have passed through the group a few times $\sim10^8$ yr ago. It has been suggested that NGC 7320C is responsible for the tidal tail of NGC 7319 \citep{moles97}. Such galaxies may thus be transient group members that nevertheless could have a significant effect on CG galaxy evolution. 

\subsection{Comparison to HCGs \label{comp_to_HCGs}}
We compare the properties of the CGs that we found in the De Lucia \& Blaizot SAM output to those of the CGs 
found via the observational work of Hickson and collaborators.
In particular, we compare the space densities, the median radii, and the median 3D galaxy velocity dispersions of the groups,
\(\sigma_{v,\rm 3D} \equiv \sqrt{\sigma_{v,x}^2 + \sigma_{v,y}^2 + \sigma_{v,z}^2}\). 
The results are summarized in Table~\ref{parameter_sets}.

The number of CGs identified in Hickson's catalog can be compared to the number of CGs identified in the Millennium Simulation galaxy catalogs for specific parameter sets by making a few assumptions. While \citet{hickson92} identified a total of 92 groups\footnote{Among these, a further four groups may be questionable; see \citet{diaz-gimenez12} and references therein.} with at least three accordant members within 
$z\le 0.14$
over 67\% of the sky, the median redshift of groups in the catalog is $z=0.03$, which suggests that the catalog becomes increasingly incomplete for larger values of $z$.  Further, \citet{diaz-gimenez10} found that the 
velocity-filtered
HCG catalog is only 8\% complete even at $z=0.03$.  
To this redshift, and over 67\%\ of the sky, there are 47 detected HCGs with accordant velocities in \citet{hickson92}.
Therefore, the expected number of CGs over 67\% of the sky out to $z=0.03$ is $N_{\rm HCG} = 47/0.08 = 587.5$, if we assume that the Hickson catalog is 8\%\ complete out to this distance. On the other hand, the comoving volume to $z=0.03$ is
$V=(\Delta \Omega/3) (D_L/(1+z))^3$,
where $D_L = 92.1 \, h^{-1}$ Mpc is the luminosity distance to this redshift
and $\Delta\Omega = 8\pi/3$ is the solid angle for a 67\%\ sky coverage.
This gives a completeness-corrected expected
number density of HCGs to $z=0.03$ of $n_{\rm HCG}=2.9\times10^{-4} \, h^3 \, {\rm Mpc}^{-3}$.

The volume of a snapshot of the Millennium Simulation is $V_{\rm snap} = (500~h^{-1}{\rm Mpc})^3 = 1.25\times 10^8 \, h^{-3} $~Mpc$^3$,
and there is one snapshot at $z=0$.
Hence, if the simulation has the same volume density of groups as HCGs, we expect the total number of identified groups in the simulation to be $N_{\rm sim} = 36,250$ at $z=0$.  
This number is between the number produced by set E and the ``lenient'' parameter set (see Table~\ref{parameter_sets}, Column (10)). Thus, if one assumes that the space density of HCGs up to $z=0.03$ is representative, a more lenient set of parameters is preferred.

The median projected group separation in the \citet{hickson92} catalog is $R_{\rm HCG}= 39 \, h^{-1}$~kpc.
Thus, if the 3D separation is $2\pi/5$ larger than the projected 2D separation, then the expected 3D HCG galaxy separation is $\sim 49\,h^{-1}$~kpc.
Several of the parameter sets produce median group radii that are within 20\% of 
this value,
including the ``default,'' C, D, E, and F parameter sets.  All of these parameter sets share a common neighborhood ({\sc nh}) of $50\, h^{-1}$ kpc, 
unlike all the rest of the parameter sets.

While there are parameter sets that reasonably reproduce the observed space density and sizes of HCGs, the 3D velocity dispersion presents an issue. \citet{hickson92} observed a 
median
1D velocity dispersion of $200$~km~s$^{-1}$ and,
after
considering the uncertainties in the individual velocities,
inferred a 3D dispersion of $\sigma_{v, \rm 3D,HCG}=331\, {\rm km \, s}^{-1}$. The parameter sets tested here, however, 
all
produce significantly smaller 3D velocity dispersions that are all $<120 \, {\rm km \, s}^{-1}$, and typically $\lesssim$100~${\rm km \, s}^{-1}$.  
We note that for groups with only three members, on average, the 3D velocity dispersion we obtain is 81 km s$^{-1}$, whereas for groups with four members or more, on average, the 3D velocity dispersion is 159 km s$^{-1}$. This shows that, even if we had set \nmin~=~4, we would still not be able to reconcile our results with the HCG value of 331 km s$^{-1}$.

We have performed a series of further tests to assess whether any specific modifications in our approach might result in substantially different velocity dispersions, but we were only able to obtain negligible changes. Specifically, we compared results from DBscan both including and excluding dwarf galaxies. We also computed 3D velocity dispersions using the biased dispersion, unbiased dispersion, and gapper techniques, resulting in median velocity dispersions at $z=0$ of $\sim 100$~km~s$^{-1}$, $\sim 125$~km~s$^{-1}$, and $\sim150$ ~km~s$^{-1}$, respectively, which fall well short of the value observed for HCGs.

We postulate that this disagreement may be due to the inherent differences between selecting observed groups based on 2D spatial projections as opposed to actual 3D information available in the simulations. One hypothesis is that this key difference has its origin in the more restrictive selection criteria in 3D space, resulting in more tightly bound groups than the ones in the \citet{hickson92} catalog.

\section{Discussion}
From the parameter sets used here to identify CGs, it is clear that the precise definition of CGs can have a significant impact on the resulting demographics.  Nevertheless, the different parameter sets do result in some broad similarities with respect to CG populations over cosmic time.  For example, all of the parameter sets tested here produce a rapid rise in the population of CGs up to a redshift of $z\sim 4-5$, after which the different populations reach peaks at $z\sim 1-3$ and then slowly decline.

It is noteworthy that the two parameter sets that best reproduce the properties of HCGs at redshifts of $z<0.03$ (``default'' and C) exhibit the peak in their populations at redshifts of $z=1.9$, which mirrors the peaks in both the cosmic star formation rate \citep[e.g.,][]{madau14} and the galaxy merger rate histories \citep[e.g.,][]{bertone09}.  Even during this ``peak'' epoch of CGs, according to the results for these parameter sets,
only up to $\sim 1$\% of nondwarf galaxies reside in CGs.  For these same parameter sets, only 4-5\% of nondwarf galaxies have been members of CGs at some point over cosmological time.  

Despite identifying parameter sets that reproduce the sizes and population of HCGs at $z<0.03$, we were not able to reproduce the median 3D velocity dispersion of HCGs.  The groups identified in the simulation have a
median 3D velocity dispersion of $\sigma_{v,\rm 3D} \lesssim 100$~km~s$^{-1}$, while HCGs have a
median 3D dispersion of $\sigma_{v,\rm 3D,HCG} \sim 331$~km~s$^{-1}$. 
For galaxies with \nmin~$=4$, \citet{diaz-gimenez12} found a higher median 1D velocity dispersion of $\sim 248$~km~s$^{-1}$ by means of mock redshift catalogs, which translates to $\sigma_{v,\rm 3D,HCG} \sim 430$~km~s$^{-1}$, also
higher than our median $\sim 159$ km s$^{-1}$ for such CGs. We note that the value of $\sim248 \ \rm km \ s^{-1}$ is very close to the reported value of $\sim 262$ and $\sim 237 \ \rm km \ s^{-1}$ for observed HCGs and 2MASS CGs, respectively, in \citet{diaz-gimenez12}.
This is particularly puzzling given that we adopted a relative velocity restriction of $v= \pm 1000$~km~s$^{-1}$, identical to that of \citet{hickson92}.  We postulate that this discrepancy may be, in part, due to HCGs being identified based on their apparent projected spatial proximity, although this seems unlikely to account for a factor of $\sim 3$ between observed and simulated groups. However, the recent work of \citet{tzanavaris2019} studying the 3D evolution of individual CGs found in simulations suggests that the velocity fields may be highly nonisotropic, and so such a possibility warrants further study. 
Alternatively, the discrepancy between simulated and observed CG velocity dispersions might instead represent a real limitation of simulations of this nature to reproduce the observed properties of galaxy systems in the low-mass group range, a mass range for which the simulations were not tuned. 

From the results presented here, it would appear that the CG environment is not prevalent in cosmological history.  Even with the most lenient parameter set tested here (which does not fully reproduce the properties of HCGs), only $\sim 16$\% of nondwarf galaxies have been members of a CG at some point in their evolution (Figure~\ref{fig:bestworst}). A major limitation of the Millennium Simulation is its mass resolution and the resulting exclusion of dwarf galaxies in this analysis.  Given that low-mass galaxies are the dominant population at all redshifts \citep[e.g.,][]{binggeli88}, this limitation is likely to have a significant impact on the statistics of CGs. The impact of excluding dwarf galaxies will be increasingly strong at higher redshifts where the relative population of dwarf galaxies approaches values of unity (see Figure~\ref{fig:dwarf_fraction}).

\section{Conclusions}

We investigate the prevalence of the CG environment over cosmological time using a Millennium Simulation galaxy catalog. The goals of this work are twofold: first, to constrain the fraction of galaxies that have ever existed in this unusual environment, and second, to determine whether there is an ``epoch'' of CGs in cosmological history during which this environment was particularly common.  To accomplish these goals, we use a number of tunable parameters to identify CGs in the simulation. The key parameters are varied over a range that is centered on ``default'' values that best represent properties of Hickson CGs in the local universe. The main conclusions are as follows: 

\begin{enumerate}
\itemsep-4pt 
\item Every set of parameters tested here produces a peak relative population of CGs in the range of $z\sim1$--3, while both of the parameter sets that best reproduce the properties of HCGs (``C'' and ``default'' in Table~\ref{parameter_sets}) result in a peak relative population of CGs at $z\sim 1.9$.

\item The fraction of nondwarf galaxies that are members of CGs at any redshift never exceeds $\sim 3.2$\%, even for the most lenient parameter set.  The best-fit parameter sets result in peak relative fractions of $\lesssim$1\%.

\item The fraction of nondwarf galaxies that have {\it ever} been members of a CG does not exceed \mbox{$\sim$16\%}.  The best-fit parameter sets indicate that this value is probably closer to $\sim$4\%.

\item The exclusion of dwarf galaxies from this analysis could have a significant impact on the values presented here in the sense that the relative fractions are lower limits. Including dwarf galaxies becomes increasingly important at higher redshifts.  

\item While the $z<0.03$ number density and median size of our default set of CGs match those of HCGs, the 3D velocity dispersions of CGs are about half the measured values of HCGs. This suggests that the CGs found in the Millennium Simulation galaxy catalogs are more tightly bound than observed HCGs.
\end{enumerate}

\acknowledgments

We thank the referee, G. Mamon, for detailed input that improved the presentation in this paper. C.D.W. thanks the University of Virginia Small Research Grants for undergraduates for their support of this project.  
T.V.W is supported by the NSF through the Grote Reber Fellowship Program administered by Associated Universities, Inc./National Radio Astronomy Observatory, the D.N. Batten Foundation Fellowship from the Jefferson Scholars Foundation, the Mars Foundation Fellowship from the Achievement Rewards for College Scientists Foundation, and the Virginia Space Grant Consortium.
K.E.J. is grateful to the David and Lucile Packard Foundation for their generous support. 
S.C.G. thanks the Natural Science and Engineering Research Council of Canada and the Ontario Early Researcher Award Program for support.
P.T. acknowledges support by NASA ADAP 14-ADAP14-0200 (PI Tzanavaris).

\begin{center}
  {\bf ORCID iDs}
\end{center}
Trey V. Wenger\\
Panayiotis Tzanavaris\\
Kelsey E. Johnson\\
\vspace{-0.6cm}

\bibliographystyle{likeapj}
\bibliography{ref.bib}

\end{document}